\documentclass[a4paper,11pt]{amsart}
\usepackage{graphicx}
\usepackage{amssymb}
\begin{document}

\hyphenation{gra-vi-ta-tio-nal re-la-ti-vi-ty Gaus-sian
re-fe-ren-ce re-la-ti-ve gra-vi-ta-tion Schwarz-schild
ac-cor-dingly gra-vi-ta-tio-nal-ly re-la-ti-vi-stic pro-du-cing
de-ri-va-ti-ve ge-ne-ral ex-pli-citly des-cri-bed ma-the-ma-ti-cal
de-si-gnan-do-si coe-ren-za pro-blem gra-vi-ta-ting geo-de-sic
per-ga-mon cos-mo-lo-gi-cal gra-vity cor-res-pon-ding
de-fi-ni-tion phy-si-ka-li-schen ma-the-ma-ti-sches ge-ra-de
Sze-keres con-si-de-red tra-vel-ling ma-ni-fold re-fe-ren-ces
geo-me-tri-cal in-su-pe-rable sup-po-sedly at-tri-bu-table
Bild-raum in-fi-ni-tely counter-ba-lan-ces iso-tro-pi-cally
pseudo-Rieman-nian cha-rac-te-ristic geo-de-sics
Koordinaten-sy-stems ne-ces-sary}

\title[On gravitational fields created by electromagnetic waves] {{\bf On the gravitational fields\\created by the electromagnetic waves}}

\author[Angelo Loinger]{Angelo Loinger}
\address{A.L. -- Dipartimento di Fisica, Universit\`a di Milano, Via
Celoria, 16 - 20133 Milano (Italy)}
\author[Tiziana Marsico]{Tiziana Marsico}
\address{T.M. -- Liceo Classico ``G. Berchet'', Via della Commenda, 26 - 20122 Milano (Italy)}
\email{angelo.loinger@mi.infn.it} \email{martiz64@libero.it}

\vskip0.50cm

\begin{abstract}
We show that the Maxwell equations describing an electromagnetic
wave are a mathematical consequence of the Einstein equations for
the same wave. This fact is significant for the problem of the
Einsteinian metrics corresponding to the electromagnetic waves.
\end{abstract}

\maketitle

\vskip1.20cm \noindent \small \textbf{Summary} -- Introduction --
\textbf{1}. On a consequence of the fact that the light-rays are
null geodesics in any spacetime manifold. -- \textbf{2}. The
Maxwell equations of an electromagnetic wave are a consequence of
the Einstein equations for the same wave. -- \textbf{2bis}. An
example. -- \textbf{3}. A result analogous to that of
sect.\textbf{2} holds in the linear version of GR. --
\textbf{3bis}, \textbf{3ter}. An example. -- \textbf{4}. A final
remark. -- Appendix.

\vskip0.80cm \noindent \small PACS 04.20 -- General relativity.
\normalsize

\vskip1.20cm \noindent \textbf{\emph{Introduction.}} -- Very
rarely this subject has been approached from a really physical
standpoint. On the contrary, there are in the literature many
interesting papers of a mathematical character about metrics which
are linked \emph{in some way} to the propagation of
electromagnetic (e.m.) waves. We recall sect. \textbf{8} of a
famous paper (1923) by Eddington \cite{1}, the paper of 1926 by
Baldwin and Jeffery \cite{2}, the paper by Bonnor of 1969
\cite{3}, and the overflowing of geometrical articles on the
plane-fronted waves with parallel propagation (briefly, p-p waves)
\cite{4}.

\par The classic treatment by Tolman \cite{5}, which is limited to
the linear version of GR and to very particular models, is not
fully satisfying, because it neglects the important role of the
equation of the characteristic surfaces of Maxwell theory.

\par A new approach to the topic is given in the present paper.

\vskip1.20cm \noindent \textbf{1.} -- In any spacetime the e.m.
rays are null \emph{geodesics}, as it is well known. Consequently,
\emph{no} undulatory, purely gravitational, and autonomous field
is created by the propagation of any e.m. wave in any spacetime
manifold. (This propagation is a ``natural'' one, like that of any
e.m. wave in a Minkowski spacetime). And clearly, \emph{no
gravitational interaction exists among the various portions of an
e.m. wave}.

\vskip1.20cm \noindent \textbf{2.} -- Einstein field equations in
a space devoid of bodies and of charges are $(c=G=1)$:

\begin{equation} \label{eq:one}
R_{jk} = -8\pi E_{jk} \quad; \quad (E_{j}^{j}=0) \quad;
\quad(j,k=0,1,2,3) \quad,
\end{equation}

if $E_{jk}$ is the e.m. energy tensor, given by

\begin{equation} \label{eq:oneprime}
4 \pi E^{jk} = -F^{j}_{\,\,m} \, F^{km} + \frac{1}{4} \, g^{jk} \,
F_{mn}\, F^{mn} \tag{1$'$} \quad,
\end{equation}

where $F_{mn}$, $(m,n=0,1,2,3)$, is the e.m. field of the
considered e.m. wave. We have from Maxwell theory:

\begin{equation} \label{eq:onesecond}
F_{mn} = \Phi_{m:\,n} - \Phi_{n:\,m} = \Phi_{m,\,n} - \Phi_{n,\,m}
\tag{1$''$}\quad,
\end{equation}

if $\Phi_{m}$ is the e.m. potential; the colon and the comma
denote respectively a covariant and an ordinary derivative.
Besides eqs. (\ref{eq:onesecond}), we have Maxwell equations

\begin{equation} \label{eq:two}
\left(F^{jk} \, \sqrt{-g}\right)_{,\,k} = 0 \quad.
\end{equation}

Eqs. (\ref{eq:one}) tell us that

\begin{equation} \label{eq:three}
E^{jk}_{\,\,\,:\,k} = 0 \quad,
\end{equation}

from which, taking into account (\ref{eq:oneprime}) and
(\ref{eq:onesecond}), we get

\begin{equation} \label{eq:four}
F^{jk}_{\,\,\,:\, k} = 0  \quad ,
\end{equation}

which coincide with eqs. (\ref{eq:two}).

\par \emph{This means that if we express} $E_{jk}$ \emph{as a function
of} $\Phi_{m}$, \emph{Einstein eqs.} (\ref{eq:one}) \emph{have
Maxwell eqs.} (\ref{eq:two}) \emph{as a a mathematical
consequence}. Gravitation has ``absorbed'' the e.m. properties of
the e.m. wave. This result has as a necessary condition that the
differential equation of the characteristic surfaces
$z(x^{0},x^{1},x^{2},x^{3})=0$, \emph{i.e.}

\begin{equation} \label{eq:five}
g^{jk} \, \frac{\partial z(x)}{\partial x^{j}} \, \frac{\partial
z(x)}{\partial x^{k}} = 0 \quad ,
\end{equation}

is the \emph{same} for both Maxwell and Einstein fields (Whittaker
and Levi-Civita).

\par At this point, it is very natural to specify the reference
frame in such a way that four components of the metric tensor
$g_{jk}$ are functionally \emph{identical} to the four components
of the e.m. potential $\Phi_{m}$, which describes our e.m. wave.

\vskip1.20cm \noindent \textbf{2bis.} -- Let us consider,
\emph{e.g.}, a continuous flow of e.m. waves described by the
following four-potential $\Phi_{m}$:

\begin{equation} \label{eq:six}
\Phi_{0} =  \Phi_{1} =\Phi_{3} =0 \quad ; \quad \Phi_{2}(t+x) = A
\sin[\omega(t+x)] \quad; \quad (t\equiv x^{0}; x\equiv x^{1})
\quad ;
\end{equation}

in a \emph{Minkowskian} spacetime, eqs. (\ref{eq:six}) represent
\emph{an ordinary plane} wave (\cite{1}). We have:

\begin{displaymath} \label{eq:seven}
\left\{ \begin{array}{l} F_{21} = \Phi_{2, \,1} = A\omega \cos[\omega(t+x)] \quad, \\ \\
F_{20} = \Phi_{2, \,0} = A\omega \cos[\omega(t+x)] \quad. \tag{7}
\end{array} \right.
\end{displaymath}

The e.m. energy tensor $E_{jk}$ reduces to (\cite{2}):

\setcounter{equation}{7}
\begin{equation}\label{eq:eigth} E_{jk} = - g^{rs}
\, F_{jr} F_{ks} \quad ,
\end{equation}

and eqs. (\ref{eq:one}) give:

\begin{equation} \label{eq:nine}
R_{jk} = 8\pi g^{rs} \, F_{jr} F_{ks} \quad ;
\end{equation}

a solution of the problem can be obtained by putting, for
instance:

\begin{equation} \label{eq:ten}
g_{00} = g_{01} = g_{03} = 0 \quad ; \quad g_{02} = A
\sin[\omega(t+x)] \quad;
\end{equation}

the other components of the metric tensor, $g_{\alpha
\beta}(t,x)$, $(\alpha , \beta =1,2,3)$, are the solutions of eqs.
(\ref{eq:nine}) in which we have substituted the values
(\ref{eq:ten}). The e.m. field of the e.m. wave is thus fully
described by its own gravitational field.

 \vskip1.20cm \noindent\textbf{3.} -- In the linear version (LV)
 of GR we have approximately:

\begin{equation} \label{eq:eleven}
g_{jk} = \eta_{jk} + h_{jk} \quad,
\end{equation}

where $\eta_{jk}$ is the Minkowski tensor $(1,-1,-1,-1)$, and the
$h_{jk}$'s are small  deviations from it. LV is a
Lorentz-invariant theory. Its equations are also invariant under
the following gauge transformation of the symmetric tensor
$h_{jk}$:

\begin{equation} \label{eq:twelve}
h_{jk} \rightarrow h_{jk} + \xi_{j, \, k} + \xi_{k, \, j} \quad ,
\end{equation}

where $\xi_{j}(x)$ is an infinitesimal vector function of
$(x^{0},x^{1},x^{2}x^{3})$. Equivalently, formula
(\ref{eq:twelve}) can be viewed as the result of a transformation
of metric (\ref{eq:eleven}) under an infinitesimal change of the
Lorentzian coordinates $x^{j}$:

\begin{equation} \label{eq:thirteen}
x^{j} \rightarrow x^{j} + \xi^{j}(x) \quad .
\end{equation}

In lieu of the exact eqs. (\ref{eq:one}), we have, as it is known:

\begin{equation} \label{eq:fourteen}
\frac{1}{2} \, \Box \, h_{jk} = -8\pi E_{jk} \quad ; \quad
h^{jk}_{\,\,\,,\,k} = 0 \quad ;
\end{equation}

in the following equations of sect. \textbf{2}, we must now
substitute $g_{jk}$ with $\eta_{jk}$, and the covariant
derivatives with the ordinary ones. Eq.(\ref{eq:five}) becomes:

\begin{equation} \label{eq:fifteen}
\eta^{jk} \, \frac{\partial z(x)}{\partial x^{j}} \,
\frac{\partial z(x)}{\partial x^{k}} = 0  \quad ;
\end{equation}

the light-rays are \emph{rectilinear} null-geodesics; the e.m.
waves and the field $h_{jk}$ are propagated in \emph{Minkowski}
spacetime.

\par Four components of tensor $h_{jk}$ can be identified with the
four components of the e.m. potential $\Phi_{m}$.

\vskip1.20cm \noindent \textbf{3bis.} -- Let us consider in the LV
a continuous flow of \emph{plane} e.m. waves, described by the
four-potential $\Phi_{m}$ of eqs. (\ref{eq:six}). We have
(\cite{1}):

\begin{equation} \label{eq:sixteen}
E_{00} =  E_{01} (=E_{10}) =E_{11} = A^{2} \omega^{2}
\cos[\omega(t+x)] \quad ;
\end{equation}

the other components of $E_{jk}$ are equal to zero.

\par Accordingly,

\begin{equation} \label{eq:seventeen}
\Box \, h_{00}(t,x) =  \Box \, h_{01}(t,x) =  \Box \, h_{11}(t,x)
= -16\pi A^{2} \omega^{2} \cos^{2}[\omega(t+x)] \quad;
\end{equation}

\begin{equation} \label{eq:seventeenprime}
\Box \, h_{jk} = 0  \tag{17$'$} \quad, \quad \textrm{for }
(j,k)\neq [(00), (01), (11)] \quad .
\end{equation}

The e.m. field does not appear in eqs. (\ref{eq:seventeenprime}),
and therefore these $h_{jk}$'s can be put equal to zero: they do
not ``feel'' the action of the e.m. waves. However, if we prefer
to follow the procedure of sect. \textbf{2bis}, we can put,
\emph{e.g.}, (with gauge transformed $h_{jk}$'s ):

\begin{equation} \label{eq:eighteen}
h_{02}(t+x) = \Phi_{2}(t+x) =  A \sin[\omega(t+x)] \quad ;
\end{equation}

\begin{equation} \label{eq:eighteenprime}
h_{03} = \Phi_{3} = 0  \tag{18$'$} \quad; \quad h_{12} = \Phi_{1}
= 0  \quad; \quad h_{13} = \Phi_{0} = 0  \quad;
\end{equation}

\begin{equation} \label{eq:eighteensecond}
h_{22}= h_{23} = h_{33} = 0  \tag{18$''$} \quad.
\end{equation}

Then, eqs. (\ref{eq:seventeen}) can be re-written as follows:

\begin{equation} \label{eq:nineteen}
\Box \, h_{00} = \Box \, h_{01} = \Box \, h_{11} = \left(
\frac{\partial h_{02}}{\partial \zeta}\right)^{2} \quad ,
\end{equation}

if $\xi :=t+x$; the e.m. field is thus fully described by its own
gravitational field.

 \vskip1.20cm \noindent \textbf{3ter.} -- One finds easily the
 solution of eqs. (\ref{eq:seventeen}). Indeed, the solution of
 d'Alembert inhomogeneous equation

\begin{equation} \label{eq:twenty}
\frac{\partial ^{2}F}{\partial t^{2}} - \frac{\partial
^{2}F}{\partial x^{2}}= -16\pi A^{2} \omega^{2}
\cos^{2}[\omega(t+x)]
\end{equation}

is given by $F(t,x)=F_{0}(t,x)+F_{1}(t,x)$, where

\begin{equation} \label{eq:twentyone}
F_{0}(t,x) = \varphi(t+x) + \chi(t-x)
\end{equation}

is the general solution of the  homogeneous equation, with
$\varphi$ and $\chi$ any functions of their arguments, and
$F_{1}(t,x)$ is given by

\begin{equation} \label{eq:twentytwo}
F_{1}(t,x) = -16\pi A^{2} \omega^{2} \cdot (t-x) \left\{
\frac{1}{2} \, (t+x) + \frac{1}{4\omega} \sin [2\pi (t+x)]\right\}
\quad .
\end{equation}

Of course, only $F_{1}(t,x)$ concerns the gravitational field
generated by our e.m. waves.

 \vskip1.20cm \noindent\textbf{4.} -- \emph{A final remark}. Hilbert \cite{6} considered
 the coupled equations of Einstein and Mie (with the gravitational
 and e.m. potentials, $g_{jk}$ and $q_{j}$, as unique dynamical
 variables) in lieu of the coupled equations of Einstein and
 Maxwell. According to Mie's theory \cite{7} (which has
 revealed itself as unpractical), the electric charges would
 emerge as solutions of the field equations for the potential
 $q_{j}$, while in Einstein's theory the point-masses emerge as
 singularities of the metric tensor $g_{jk}$. Hilbert remarked
 that Mie's equations, referred to the general-relativistic
 metric, are an analytical \emph{consequence} of Einstein's
 equations with Mie's e.m. energy tensor. Therefore, for the $14$
 components of the potentials $g_{jk}$ and $q_{j}$, we have only
 the $10$ functionally independent Einstein's equations.

 \par In our previous treatment of the gravitational field created
 by an e.m. wave we had a formalism which is quiete analogous,
 from the \emph{mathematical} standpoint, to the above Hilbertian
 formalism, with Maxwell's e.m. potential $\Phi$; instead of
 $q_{j}$. We have applied and developed Hilbert's remark.

\vskip2.00cm
\begin{center}
\noindent \small \emph{\textbf{APPENDIX}}
\end{center} \normalsize

\vskip0.40cm \noindent Our approach is a \emph{direct} one (like
that of Tolman for the linear version of GR (\cite{5}): we start
from a \emph{given} e.m.--wave potential and give a prescription
to compute the generated gravitational field according to Einstein
field equations. On the contrary, we find in the literature a
clear prevalence of an \emph{indirect} method: one postulates
intuitively the more or less detailed structure of a
$\textrm{d}s^{2}$ -- or of a gravitational potential $g_{jk}$;
then, from the corresponding expression for $R_{jk}-(1/2)g_{jk}R$,
one derives the tensor $T_{jk}$ and one verifies whether it can be
interpreted as the energy tensor $E_{jk}$ of an e.m. wave. This
method has a weak point: it depends on an interpretation, and thus
its physical meaning can be dubious. This adjective is appropriate
also for those mixed procedures that make a partial use of both
the above mentioned methods.

\end{document}